\documentclass[prx,floatfix,pdftex,twocolumn,ltxgrid]{revtex4}

\usepackage[pdftex]{graphicx}
\usepackage{subfig}
\usepackage{stdclsdv}
\usepackage{array}
\usepackage{amsmath}
\usepackage[version=3]{mhchem}

\usepackage{hyperref}
\usepackage{enumitem}

\newcommand{\av}[1]{\langle #1 \rangle}

\begin{document}

\title{ Favoured Local Structures in Liquids and Solids: a 3D Lattice Model}

  \author{Pierre Ronceray} \affiliation{Laboratoire de Physique
Th\'eorique et Mod\`eles Statistiques, Univ. Paris-Sud,
B\^at. 100, 91405 Orsay Cedex, France}

\author{and Peter Harrowell} \affiliation{School of Chemistry,
University of Sydney, Sydney N.S.W. 2006, Australia}

\begin{abstract}
  We investigate the connection between the geometry of Favoured Local
  Structures (FLS) in liquids and the associated liquid and solid
  properties. We introduce a lattice spin model -- the FLS model on a
  face-centered cubic lattice -- where this geometry can be
  arbitrarily chosen among a discrete set of $115$ possible FLS. We
  find crystalline groundstates for all choices of a single
  FLS. Sampling all possible FLS's, we identify the following trends:
  i) low symmetry FLS's produce larger crystal unit cells but not
  necessarily higher energy groundstates, ii) chiral FLS's exhibit to
  peculiarly poor packing properties, iii) accumulation of FLS's in
  supercooled liquids is linked to large crystal unit cells, and iv)
  low symmetry FLS's tend to find metastable structures on cooling.

\end{abstract}

\maketitle

\section{Introduction}
\label{sec:intro}

For a given interaction potential between particles, some local
arrangements will be particularly stable and these \emph{favoured
  local structures} (FLS) will thus accumulate in the liquid upon
cooling. There is a considerable body of literature on the geometric
and kinetic consequences of these FLS's in supercooled liquids, which
have been the subject of a number of recent
reviews~\cite{ma-review,royall-review}. A possibility that has
remained of perennial interest since first suggested by Frank in
1952~\cite{frank} is that if a single FLS dominates the liquid
structure and its geometry is incompatible with crystalline
symmetries, it could stabilize the liquid down to the point of
dynamical arrest.  This is the starting point of the \emph{geometrical
  frustration} approach to understanding the glass
transition~\cite{tarjus-review}. The fundamental question of this
field can be posed as follows: might there be favoured local
structures of such geometrical awkwardness that they either do not
have a periodic ground state or, if they do, it is so high in energy
that the freezing point is forced down well below the limit of kinetic
accessibility?

To properly address this question it is important that we consider the
consequences of any particular set of FLS's for \emph{both} the
crystal phase as well as the liquid state. Indeed, the stability of
the liquid depends crucially on the stability of the crystal and the
kinetics of transformation into that ordered structure. In adopting
this perspective, physical consistency requires that the crystal
structure must be assembled out of the same set of FLS's available to
the liquid. The fact that the simulation studies of liquid FLS's
typically report little or no concentration of the crystalline local
structure~\cite{ma-review,royall-review} (despite their stability) is,
we suggest, a consequence of the high symmetry of the local
environments found in the small unit cell crystals frequently
represented in simulation models. As we have previously
shown~\cite{ronceray2}, a liquid can only accommodate a low
concentration of high symmetry FLS's due to their high entropy
cost. In cases where the crystal unit cell is larger than just the
nearest neighbour coordination, the crystal local structure has been
observed at high densities in the supercooled liquid~\cite{ulf}.

Reports of systematic searches of low energy or high density crystal
structures have appeared with increasing frequency in recent
years~\cite{crystals}. These studies have highlighted the
extraordinary variety of crystal structures that can be accessed by
the parametrization of simple shapes. The role of local structure in
liquids remains a more challenging question. The structure of molten
salts has generated an extensive literature~\cite{salts} as has the
structures in liquids characterized by local tetrahedral order such as
silica~\cite{silica}. Even in cases when the local structures are
complex and multiple, there exists a strong conviction that they
represent an essential key to understanding the properties of the low
temperature phases. A considerable body of work addresses the favoured
local structures in binary and ternary atomic
alloys~\cite{ma-review}. These local structures have been associated
with dynamic arrest and slow crystallization rates~\cite{ma-review}.

Our objective is to complete a systematic study of how changes in the
geometry of the FLS influence the properties of the liquid and crystal
phases. To carry this out through adjustments of the interparticle
potential would be difficult, requiring the invention of a set of
complex potentials for the task. In this article we shall circumvent
this problem by explicitly defining our Hamiltonian in terms of the
FLS's themselves. In this way we simply assign potential energies to
each of the possible FLS's so as to favour selected local
structures. As the number of possible FLS's can be large, even this
simple recipe can generate an enormous number of possible Hamiltonians
so we shall restrict our attention to the cases where only a single
local structure at a time is favoured. With this restriction we shall
not be able to directly address the original Frank
picture~\cite{frank,tarjus-review} which requires at least two FLS's -
an icosahedral structure assumed to dominate the liquid structure and
the cube octahedron that characterises the local structure of the face
centred cubic (fcc) crystal. The influence of multiple FLS's we leave
for future work.

In this paper we implement this FLS Hamiltonian using Ising spins on
the vertices of an fcc lattice. We have previously reported on the
features of this FLS model in two dimensions, where we explored the
relationship between the FLS and the crystal
structure~\cite{ronceray1}, liquid entropy~\cite{ronceray2}, the
freezing transition~\cite{ronceray3} and the role of
chirality~\cite{ronceray4}.  In this paper we shall examine how these
various observations are influenced by the increase to 3D. The
immediate challenge of the 3D FLS model is the significant increase in
the number of possible FLS's - from $9$ in 2D to $115$ in $3D$.

The paper is organised as follows. In Section~\ref{sec:model}, we
detail the definition of the three-dimensional model. We then turn to
the description of the different regimes of interest: in
Section~\ref{sec:GS}, we investigate the structural properties of the
ground states, finding that they are all crystalline but can be very
complex. In Section~\ref{sec:thermal}, we describe the thermal
behaviour of the FLS model: liquid phase, supercooling and freezing
transitions. Finally, Section~\ref{sec:stats} is devoted to an
analytic study of the previously studied quantities, using statistical
methods to connect structural and thermal properties. We discuss the
consequences of our findings in Section~\ref{sec:discussion}.  Details
of the groundstate properties and transition behaviour for each FLS is
provided as Supplementary Material~\cite{supp}.

\section{The Model}
\label{sec:model}

\begin{figure*}%
    \begin{center}
    \includegraphics[width=0.9\textwidth]{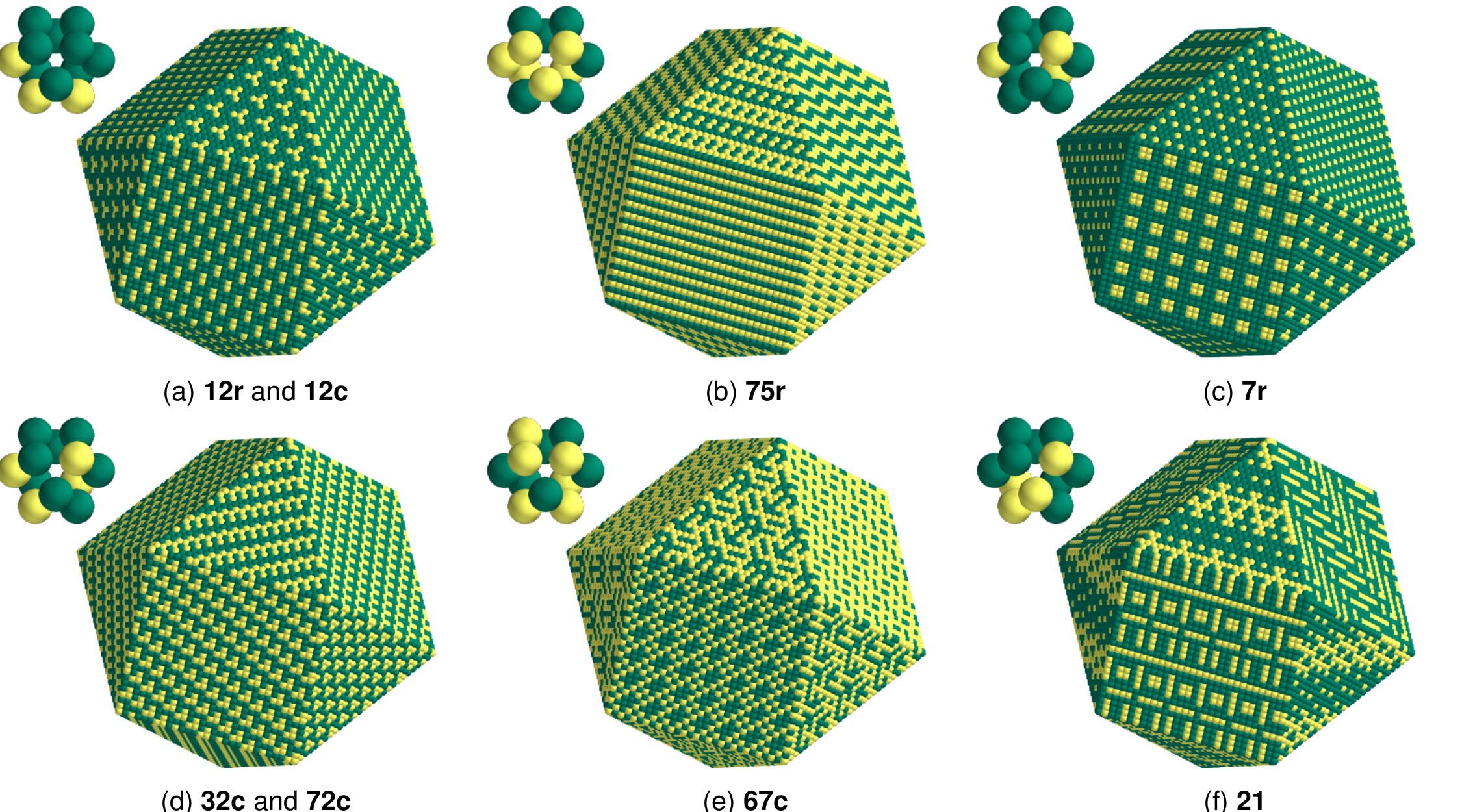}
    \caption{ A selection of ground state structures for various
      choices of the favoured local structure. This representation
      shows a portion of an infinite crystal, cut along different
      planes. The associated FLS is depicted in inset, while the
      labels correspond to those employed in Supplementary
      Material. For chiral FLS's, the letters \textbf{c} and \textbf{r}
      respectively correspond to systems with one and both enantiomers
      favoured.  }
    \label{fig:GS_pictures}
  \end{center}
\end{figure*}

The Favoured Local Structures model is a model of Ising spins
(\emph{i.e. }binary variables) located at the vertices of a regular
lattice. In this article we focus on the case of a three-dimensional
face-centered cubic (fcc) lattice. The local structure at a given
lattice site is defined as the spin conformation of the nearest
neighbours of this site (regardless of the value of the spin at said
site). Thus there are $2^z$ possible local structures at a given site,
where $z=12$ is the connectivity of the lattice. Among this large
number of structures, some differ only by a rotation and are treated
as indistinguishable, so that this model is isotropic.  An instance of
the model is obtained by selecting one local structure and making it
\emph{favoured} by directly stabilizing it. In order to do so, we
attribute an energy of $-1$ to sites whose local environment
corresponds to this \emph{favoured local structure} (up to a
rotation), while all other sites have zero energy. The total energy of
a spin configuration is thus simply, up to a minus sign, the number of
sites lying in the FLS. We consider configurations of such systems on
a large lattice with periodic boundary conditions, in the canonical
ensemble. The temperature thus controls the density of Favoured Local
Structures, and the zero-temperature limit corresponds to dense
packing of FLS on the lattice.

On the two-dimensional triangular lattice previously
studied~\cite{ronceray1,ronceray2,ronceray3,ronceray4}, $z=6$ and thus
there were $64$ possible local structures, falling into $14$
rotationally distinct classes. As the model turns out to be the same
when one FLS or its spin inverted variant is selected, there were $9$
possible choices left. Finally, two of these local structure were
\emph{chiral}, \emph{i.e.} they lacked a reflection symmetry, and
formed an \emph{enantiomer pair}. While the model would be identical
by selecting either of those enantiomers, we found interesting to
distinguish the case when only a one of them is favoured  -- the
\emph{chiral} case -- from that when both are favoured -- the
\emph{racemic} case. Physically, the racemic case might correspond
either to an equimolar mixture of enantiomers or to a flexible
molecule with two enantiomeric conformations. This left us with $9$
possible choices for the FLS, all of which were studied in detail.

In this article we consider the case of a face-centered cubic lattice,
for which each site has $z=12$ neighbours. The $4096$ local structures
reduce to $115$ after removing those which are related by rotation or
spin inversion of another. Among them, there are $39$ enantiomer pairs
for which we distinguish the racemic and the chiral cases, while the
remaining $37$ structures are achiral. A selection of these
structures, where spins are represented by coloured spheres, is
presented in the insets of Figure~\ref{fig:GS_pictures}. All these
structures are depicted, labelled and individually studied in the
Supplementary Material~\cite{supp}.  We now report on the study of
these $115$ possible systems.

The main question we want to address is the influence of the local
geometry on the macroscopic properties of the system. To quantify the
degree of symmetry of the local structure, we introduce the size
$\mathcal{S}$ of the spatial point group of an FLS. This number ranges
from $\mathcal{S}=1$ for those FLS's with no symmetry save the
identity operation to $\mathcal{S}=48$, the size of the complete
symmetry group of the lattice. The latter is attained only by one FLS,
the all-up environment labelled \textbf{0}, corresponding to
ferromagnetic ordering. The inclusion of the racemic mixtures
complicates the test for the influence of $\mathcal{S}$, since the
added flexibility due to favouring two distinct enantiomers is not
reflected by this symmetry parameter. Besides, we show in
Section~\ref{sec:chiral_special} that the chiral symmetry is peculiar
in our model. For these reason, we distinguish the sets of achiral,
chiral and racemic FLS throughout the paper.

\section{Structural properties and ground states}
\label{sec:GS}

In this Section, we present the ground state properties of the model,
corresponding to dense packing of the FLS on the lattice.

\subsection{Ground state energies}
\label{sec:Eo}

The ground state energies $E_o$, corresponding to minus the maximum
packing density of the FLS, have been determined for all choices of
FLS. These energies and the associated crystalline structures were
determined by an enumerative process similar to that introduced by
Hart and Forcade~\cite{hart}, detailed in the Appendix. \emph{We found
  no evidence of a non-crystalline structure more stable than the
  crystals identified by this method.} A selection of crystalline
structures is presented in Figure~\ref{fig:GS_pictures}. A complete
presentation of these structures is provided in ref.~\cite{supp}.

\begin{figure}[t]
    \begin{center}
     \includegraphics[width=\columnwidth]{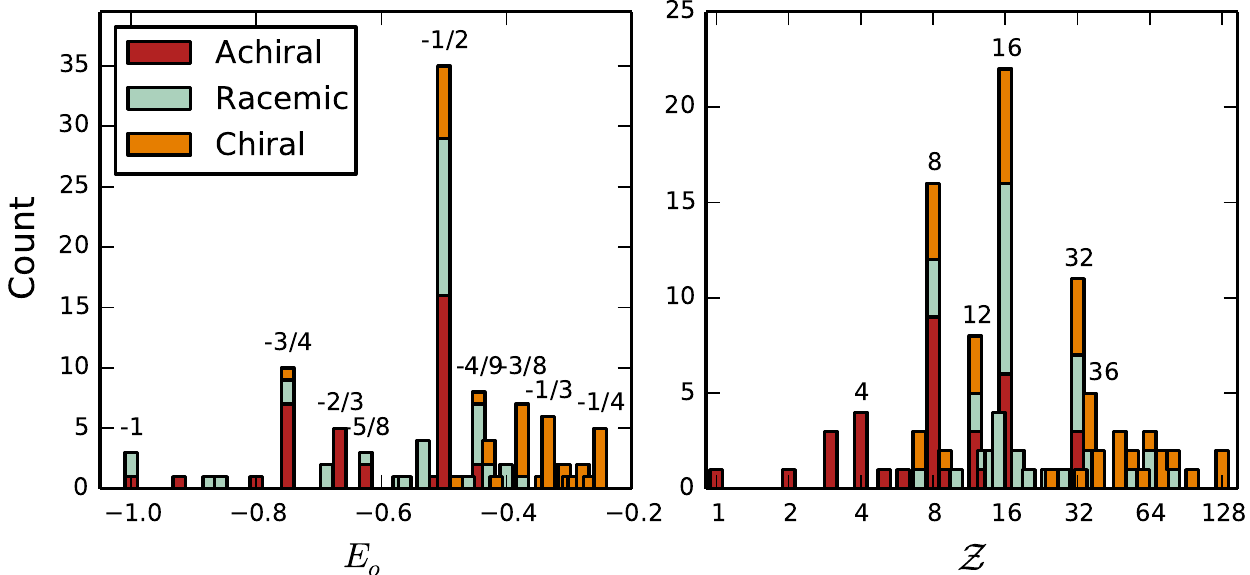}
     \caption{\label{fig:Eo_and_Z}The distribution of ground state energies
       $E_o$ (\emph{left panel}) and crystal cell size (\emph{right
         panel}). The colors indicate the nature of the FLS.  }
  \end{center}
\end{figure}

The distribution of ground state energies is plotted in
Figure~\ref{fig:Eo_and_Z}. These energies are integer fractions,
corresponding to minus the ratio of the number of FLS in the crystal
cell to the size $\mathcal{Z}$ of this crystal. The ground state
energies range from $-1$ (corresponding to all sites in the FLS,
attained by three local structures) to $-1/4$, the lowest packing
fraction obtained (attained by five chiral FLS's). Notably, chiral
FLS's tend to have higher ground state energy than other structures, a
fact that will be discussed in Section~\ref{sec:chiral_special}. There
is a particularly large peak at $E_o = -1/2$ characterized mostly by
structures consist of an alternating 2D layers that are either
completely filled or completely empty of FLS.

With an average of $\av{E_o} = -0.51 (0.17)$~\footnote{ The number in
  parentheses correspond to the standard deviation of the
  distribution, over the considered set of FLS.}, the minimal energies
are generally considerably higher than those observed in
2D~\cite{ronceray1}, for which $\av{E_o} = -0.73(0.13)$. This effect
of dimension is analogous to the well documented decrease in maximum
packing density for hard spheres with increasing
dimensionality~\cite{highD}. The increase in the number of nearest
neighbour sites resulting from an increase in spatial dimension
appears to generate more constraints with regards to packing than it
provides degrees of freedom to resolve them, giving rise to the
observed decrease in packing density.

All but one of the racemic groundstates, for which both enantiomers
are equally favoured, consist of actually racemic crystals with equal
numbers of each enantiomer. This outcome is not imposed by the
Hamiltonian, as evidenced by the one exception, the racemic FLS
$\mathbf{12r}$, whose groundstate contains only a single enantiomer,
resulting in a crystal structure identical to that of the chiral
option $\mathbf{12c}$.  This crystal is depicted in
Figure~\ref{fig:GS_pictures}(a). In the case of chiral organic
molecules, there is considerable interest in understanding the
conditions under which crystallization might select a single
enantiomer from a racemic mixture. Starting with Pasteur's famous
separation of the enantiomers of tartaric acid~\cite{pasteur},
enantiomer-selective crystallization represents an important route to
obtaining single enantiomer populations of molecules. Empirically,
chiral organic molecules are found to crystallize preferentially into
the pure enantiomer in $10\%$ of cases~\cite{chiral}. Our one example,
the FLS $\mathbf{12r}$, is insufficient to establish any general
rule. That said, we note the groundstate energy of the chiral crystal
$\mathbf{12c}$, $E_o = -3/4$, is only case of a chiral crystal with an
energy lower than $-1/2$.

Our FLS model clearly demonstrates that the racemic crystal is
generally more stable than its chiral counterpart. This stability is
highlighted by the fact that, of the three choices of FLS which
achieve perfect packing, i.e. $E_{o}=-1$, two are racemic;
$\mathbf{74r}$ and $\mathbf{75r}$ (see Figure~\ref{fig:GS_pictures}(b)
for the structure of the latter case). These perfect packings, along
with the general trend for enantiomers to pack in equal proportion
point to a general packing benefit that is afforded low symmetry FLS's
when they can be paired with their enantiomer. The resulting dimer has
a higher symmetry (gaining, at the least, a centre of inversion) and
this allows for a greater number of crystal types to be considered
which, in turn, would be expected to lead to a lower groundstate
energy.

\subsection{Crystalline cell sizes}
\label{sec:Z}

The unit cell size $\mathcal{Z}$ of the groundstate crystal structure
provides some measure of the structural complexity of the crystal
since the larger $\mathcal{Z}$, the number of lattice sites within the
unit cell, the greater the information required to specify the
structure. In Figure~\ref{fig:Eo_and_Z} we plot the distribution of unit cell
sizes, revealing a large variation, ranging from $1$ to $128$, with
peaks at powers of $2$. A selection of crystalline ground state
structures is presented in Figure~\ref{fig:GS_pictures}.

The size of the largest unit cells observed in the FLS model is of
particular interest. On a practical point, a unit cell of 128 sites
highlights the diffiiculty of ensuring that the simulation volume is
appropriately shaped to accommodate the crystal. The occurrence of
crystals with large unit cells is also a topic of considerable
physical interest. Pauling, in 1955, solved the structure of an
intermetallic alloys, \ce{NaCd_2}, to find that the unit cell was
huge, containing $1152$ atoms~\cite{pauling}. Since this pioneering
result, intermetallic alloys have provided a growing number of what
have been labelled \emph{ giant unit cell} structures~\cite{urban},
with hundreds of cases of cells whose size exceeds $10^3$ atoms. The
study of such giant unit cell structures has been
identified~\cite{urban} as "one of the last white spots on the map of
metal physics". At issue are the twin problems of the description of
their organization and of the identification of the mechanism by which
they are formed from the liquid state.

We find two chiral FLS's, \textbf{28c} and \textbf{67c}, which exhibit
groundstates with a cell size $\mathcal{Z} = 128$. The groundstate of
the the FLS \textbf{67c} is shown in
Figure~\ref{fig:GS_pictures}(e). Despite this apparent structural
complexity, we find both crystals nucleate and grow spontaneously in
Monte-Carlo simulated annealing of sufficiently large systems,
confirming that they are indeed kinetically accessible from the
supercooled liquid. A preliminary examination of the crystal
structures of the \textbf{28c} and \textbf{67c} groundstates indicate
the presence of asymmetric unit cells, $8 \times 4 \times 4$ and $8 \times 8 \times 2$
respectively, characterised by repeated motifs of clusters of
FLS's. Unravelling the energetics and kinetics of assembly of such
hierarchical structures remains an interesting challenge.

\section{Thermal properties}
\label{sec:thermal}

We now turn to the study of the FLS model at finite temperature in the
canonical ensemble. We first present the properties of the equilibrium
liquids and the equilibrium freezing transitions. We then move to the
non-equilibrium properties, by investigating the dynamics in the
supercooled liquid and the low-temperature fate of an annealed system.

\begin{figure*}%
    \begin{center}
    \includegraphics[width=\textwidth]{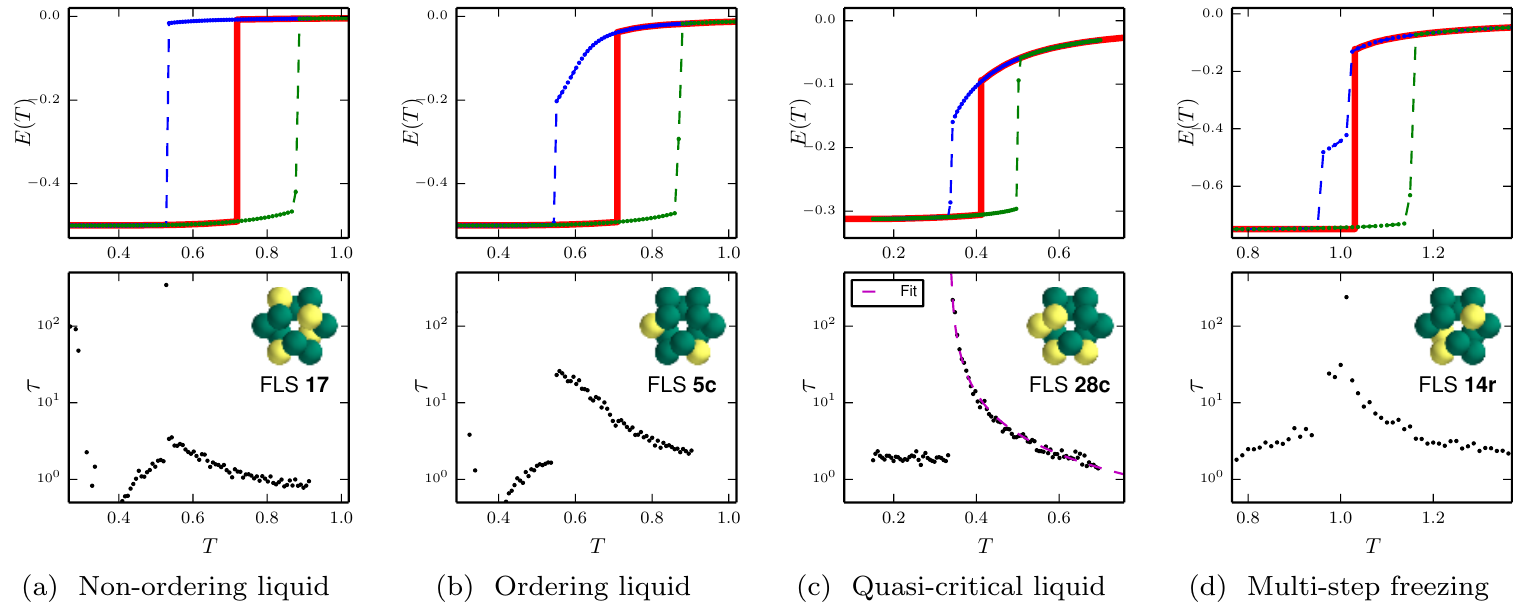}
    \caption{ Top row: The energy \emph{versus} temperature
      curves for a selection of FLS's. Monte-Carlo simulations of an
      hysteresis cycle are shown in blue and green dots (respectively
      cooling and heating), with the dashed line indicating freezing
      and melting transitions. The thick red curve corresponds to a
      thermodynamic calculation of the equilibrium curve by a Maxwell
      construction. Bottom row: the energy autocorrelation time,
      in log-scale.}
    \label{fig:E_vs_T}
  \end{center}
\end{figure*}

\subsection{Equilibrium liquids and freezing transitions}
\label{sec:liquids}

Top panels of Figure~\ref{fig:E_vs_T} present the temperature
dependence of the average energy (\emph{i.e.}, minus the FLS density)
in canonical ensemble simulations of the model, for various choices of
the FLS. A striking feature of these curves is the strongly
first-order nature of the freezing transitions, which is generic of
the model. In an hysteresis cycle, we observe both strongly
supercooled liquid and overheated crystal. In order to locate the
\emph{equilibrium} freezing point, we employed a Maxwell construction,
as in the 2D model~\cite{ronceray3}. The resulting temperature is
indicated as a thick red vertical line in Figure~\ref{fig:E_vs_T}.

\subsubsection{Order and its Entropy Cost in the Liquids}
\begin{figure}[ht]
    \begin{center}
    \includegraphics[width=6.cm]{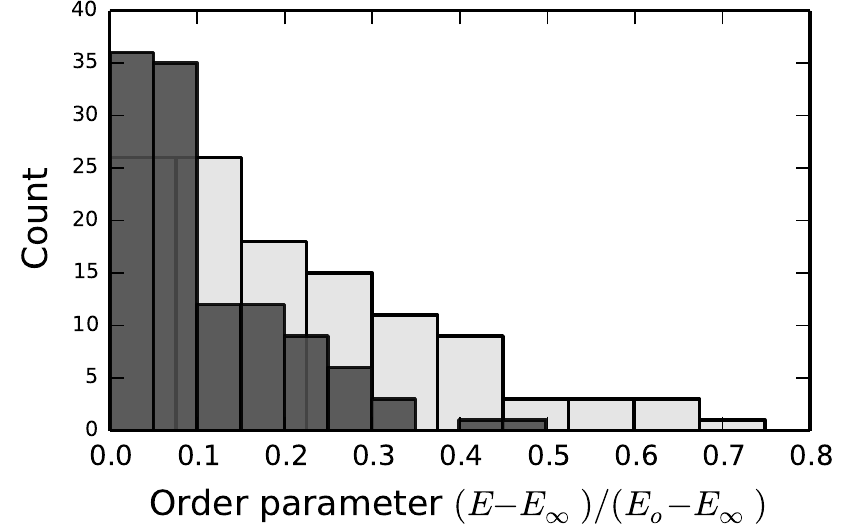}
    \caption{ Statistics for the amount of ordering. Dark gray: at the
      equilibrium crystallization transition, and light gray: at the
      limit of stability of the supercooled liquid. }
    \label{fig:order_param}
  \end{center}
\end{figure}

At infinite temperature, the liquid has a finite energy which we
denote $E_\infty$, due to random appearances of the local structure in
a completely disordered spin system. These infinite temperature FLS
densities do not exceed $0.012$. We can evaluate the ordering of the
liquid by relating its FLS density to the asymptotic regimes $T=0$ and
$T=\infty$. We introduce the order parameter $\phi$, defined
as~\cite{ronceray1}
\begin{equation}
\phi(T) =(E(T)-E_\infty)/(E_o-E_\infty)\label{eq:order_param}
\end{equation}
In Figure~\ref{fig:order_param}, we present the distribution of the
maximum value of $\phi$ attained in the equilibrium liquid
(\emph{i.e.} the value at the freezing point $T_f$). We find that it
is very small in most cases, with an average of
$\av{\phi(T_f)}=0.11(0.1)$, indicating negligible ordering in the
liquid. It is less than $0.1$ in $71$ cases out of $115$, and more
than $0.3$ in only five cases. This is in sharp contrast with the 2D
model, for which equilibrium liquids with $\phi(T_f) > 0.8$ were
observed.

The energetic gain of
increasing the density of favoured local structures competes, at
finite temperature, with the entropy loss associated to this increased
order. This entropy cost can be assessed by performing a high
temperature expansion of the energy-entropy relation, as already
discussed in the 2D case~\cite{ronceray2},
  \begin{equation}
  S = S_\infty - \frac{A}{2} (E_\infty - E)^2 + O\left((E_\infty - E)^3\right)\label{eq:SvE}
\end{equation}
where $S_\infty = \ln(2)$ (the maximum entropy of a binary spins
system), and $A$ is the entropy cost parameter quantifying the
decrease in entropy with decreasing energy.  The larger the value of
$A$, the more rapidly the liquid loses entropy as the density of FLS's
increases and the higher the temperature at which the liquid becomes
thermodynamically unstable with respect to the solid.  We established
in \cite{ronceray3} that this high temperature expansion is actually a
cluster expansion, where the coefficient $A$ is related to the
statistics of pairwise overlaps between neighbouring FLS's. We
calculated the values of A for the 3D FLS model and find them to be of
the order of $150$, considerably larger than the analogous values for
the 2D model (which are of the order of $5$). Thus the entropy cost
for a specific FLS in 3D is considerably higher than that found in
2D. This result is due to the much larger number of local structural
possibilities in 3D as compared to 2D and, hence, the greater entropy
loss when a single FLS is selected.

This high entropy cost places a serious limit on the density of FLS's
the liquid state can physically accumulate. Kauzmann~\cite{kauzmann}
noted that extrapolations of the liquid entropy to low temperatures
suggested that the entropy would vanish at some non-zero
temperature. Equation~\ref{eq:SvE}  implies an analogous disappearance
of the liquid entropy (and, hence, the 'end' of any liquid state) at
an energy $E_K$, related to $A$ by
\begin{equation}
  \label{eq:Ek}
  E_K = E_\infty - \sqrt{\frac{2S_\infty}{A}}
\end{equation}
Based on this , admittedly simplistic, estimate, the values of $E_K$
range between $-0.16$ and $-0.02$, underlining the very limited amount
of structure the 3D FLS liquids can accumulate. As discussed below,
some of the supercooled liquids in the FSL model achieve energies much
lower than $-0.16$ indicating the limits of this high-temperature
estimate, which neglects cooperative effects of more than $2$ FLS.

\subsubsection{Equilibrium Freezing Temperatures}
\begin{figure}[th]
    \begin{center}
    \includegraphics[width=0.7 \columnwidth]{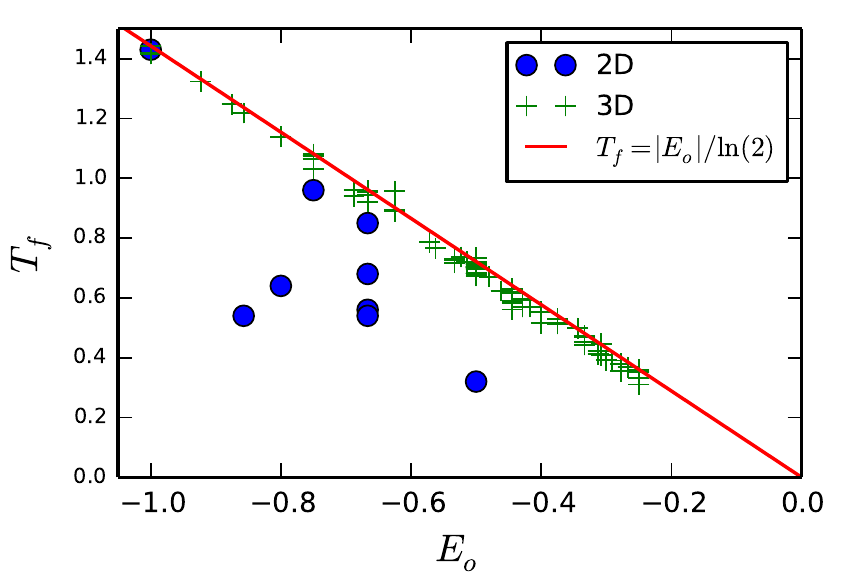}
    \caption{ The freezing temperature $T_f$ against the ground state
      energy $E_o$, for both the 3D (crosses) and the 2D (circles) FLS
      model. The line corresponds to theoretical
      Equation~\ref{eq:tf-apprx}.}
    \label{fig:freezing}
  \end{center}
\end{figure}

Given the low densities of the FLS in the equilibrium liquid, it is
reasonable to approximate the liquid with its high temperature
limit. Likewise, the crystal shows little decrease in order prior to
melting, as shown in Figure~\ref{fig:E_vs_T}. Neglecting order in the
liquid and loss of order in the crystal, we obtain an approximate
formula for freezing temperatures:
\begin{equation}
  T_f \approx \frac{|E_{o}|}{\ln 2}
  \label{eq:tf-apprx}
\end{equation}
This expression depends solely on the ground state energy, with no
dependency on the liquid properties. In Figure~\ref{fig:freezing}, we
plot the freezing temperatures obtained by Maxwell construction
against the ground state energies and compare them with the
approximate expression of Equation~\ref{eq:tf-apprx}. The data for the
2D model are also shown in the same graph.  We find that
Equation~\ref{eq:tf-apprx} provides excellent agreement with the
values of the freezing temperature for the FLS model on the FCC
lattice.  This success is in stark contrast to the 2D case where the
theoretical expression substantially overestimates $T_f$.

\subsection{Supercooling the Liquid}
\label{sec:supercooling}

Many of the FLS's result in liquids that exhibit a marked supercooling
before freezing. In this Section and the next, we employ a standard
cooling rate of $5.10^6$ rejection-free MC steps per site and per unit
of temperature, for system sizes of order $24^3$.

\subsubsection{ General Characteristics of the Supercooled Liquid}
We find that the average ratio of the minimum temperature
$T_\text{min}$ to which the liquid can be cooled to that of the
freezing point to be $\av{T_\text{min}/T_f} = 0.79 (0.09)$. This
supercooling is significant and considerably larger than observed in
the 2D model~\cite{ronceray3}, apart from the one chiral
case~\cite{ronceray4}. In molecular liquids, the glass transition
$T_g$ is typically associated with a deeper supercooling,
i.e. $T_{g}/T_{f} ~ 0.66$~\cite{2thirds}.

Supercooling is associated with an increase in the density of FLS
above that observed at coexistence. In Figure~\ref{fig:order_param},
we report the distribution of $\phi$ calculated at the lowest value of
the temperature to which each liquid can be supercooled. Here we find
a handful of liquids reaching over half of the order of the
groundstate crystal.  The average is $\av{\phi(T_{min})} = 0.21 (0.17)$,
twice that at the equilibrium crystallization transition.

\subsubsection{ Supercooling Scenarios}
In Figure~\ref{fig:E_vs_T} we plot the average energy $E(T)$ and the
energy autocorrelation time $\tau(T)$ (measured in Monte-Carlo steps
per site), estimated by the method of batch means~\cite{batchmeans} as
a function of T. We have distinguished several different scenarios
based on the behaviour of the temperature dependence of $E$ and $\tau$
in the supercooled regime. Examples of each scenario are
presented. The properties of each scenario are described as follows.

i) \emph{Non-Ordering Liquid} (79 FLS's). The most common scenario in
this model is characterized by a liquid showing negligible
cooperativity down to the freezing transition, as illustrated on
Figure~\ref{fig:E_vs_T}a. Even at the lowest temperature $T_\text{min}$ at
which the metastable liquid can be observed, the energy
autocorrelation time remains small, only $3.5$ MC steps per site.

ii)\emph{Ordering Liquid} (22 FLS's). The cases not included within
the 'Non-Ordering' category exhibit a significant accumulation of
FLS. Figure~\ref{fig:E_vs_T}b is a good example of this scenario. While
the \emph{equilibrium} liquid exhibits almost no ordering ($\phi(T_f)
= 0.07$), the supercooled liquid is characterized by a steady
accumulation of FLS, with a high and approximately constant heat
capacity $C_v = \partial E/\partial T$. In this regime, the dynamical
time scale grows steadily up to $\tau \approx 30$ MC steps per site,
indicating the existence of cooperative effects in the supercooled
liquid.

iii) \emph{Quasi-Critical Liquid} (14 FLS's). In
a number of cases, the relaxation time $\tau$ is strongly
non-Arrhenius and can climb up to $10^3$ MC steps per site (our limit
of time resolution for $\tau$), exhibiting an apparent divergence at
the end of the supercooled branch, with a correspondingly strong
increase in heat capacity. This behaviour is illustrated on
Figure~\ref{fig:E_vs_T}c. The autocorrelation time is well fitted by a
diverging power law $\tau \propto |T-T_c|^{-\gamma}$ with an exponent
$\gamma \approx 1.3$ (dashed line in Figure~\ref{fig:E_vs_T}c). While this
scenario resembles that of a glass transition, we note that this
supercooled branch terminates with a discontinuous transition in the
energy in all cases in this model.

\subsection{Low Temperature States Following Quenching}
\label{sec:lowT}

While we always observe a discontinuity in energy at the freezing
transition when annealing the system, the resulting low-temperature
state varies and the ground state crystal is not always obtained. We
now discuss the different classes of $T=0$ structure observed
following a quenching via Monte Carlo dynamics.

\begin{figure}%
  \begin{center}
  \includegraphics[width=1.05\columnwidth]{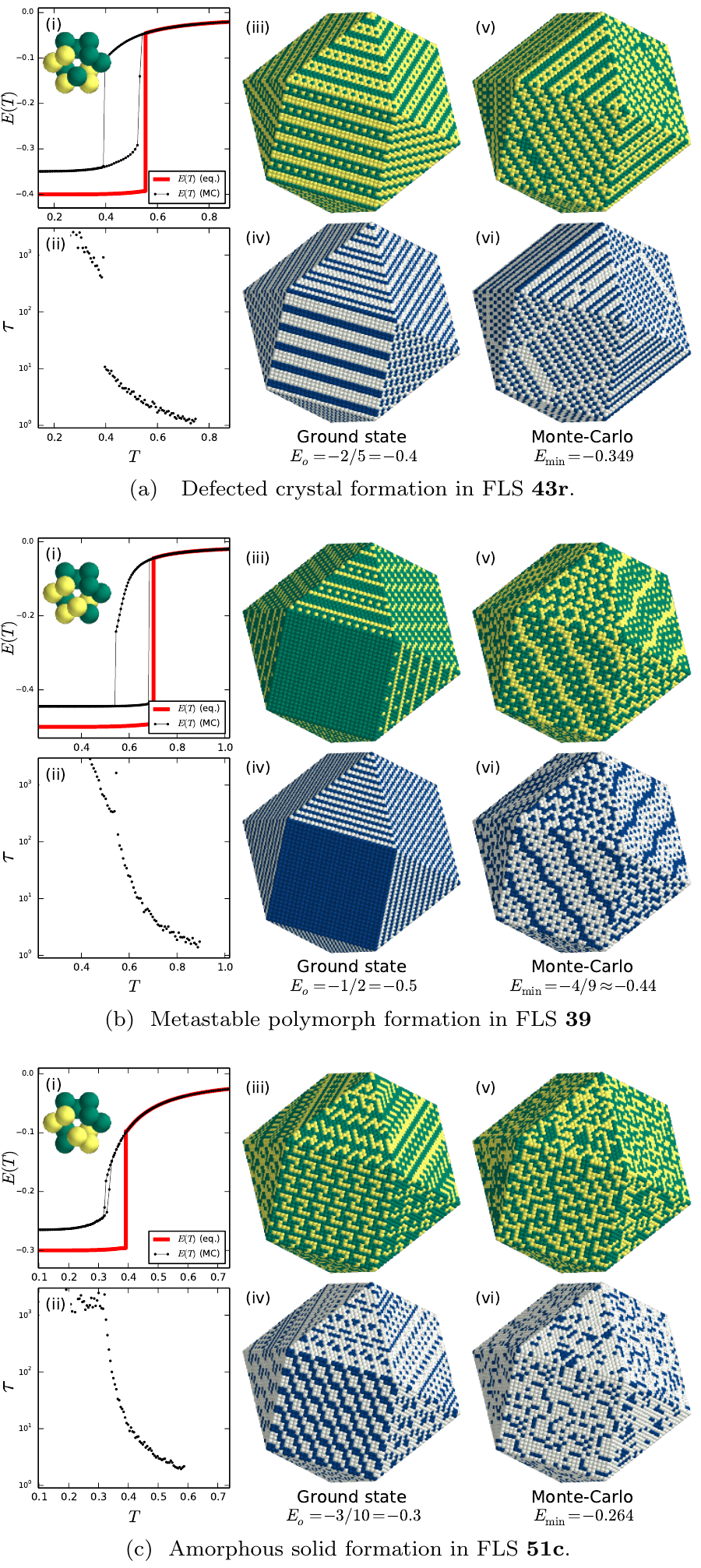} 
  \caption{ Three scenarios for avoided crystallization. In each case,
    we display: (i) the $E(T)$ curve with a picture of the FLS in
    inset and (ii) the energy autocorrelation time $\tau(T)$, as in
    Fig.~\ref{fig:E_vs_T}; (iii) the computed crystalline structure
    and (v) a low-temperature state from Monte-Carlo simulations; (iv)
    and (vi) the respective associated location of the favoured local
    structures (FLS sites in blue, non-FLS in white).  }
    \label{fig:low_T}
  \end{center}
\end{figure}

i) \emph{ Standard Crystallization} (99 FLS) The most common scenario
corresponds to the formation of a crystal structurally similar to the
ground state computed from the enumerative method. There is
some variability between cases in this family. For 66 FLS's, the
observed state corresponds exactly to the computed ground state,
provided that the boundary conditions were adapted to the crystal
structure. In 9 FLS's, we find the predicted groundstate crystal is
formed but as polycrystalline, with domains separated by grain
boundaries, as illustrated in Figure~\ref{fig:low_T}a. In a rather large
number of cases (24 FLS's), the system cooperatively orders into a
state with the same energy as the calculated ground state, but with
apparent degrees of freedom remaining. We tentatively qualified these
states as ``plastic'', without intending to assert any specific
connection with the more clearly defined plastic crystal phases of
molecular systems~\cite{plastic}. This is reminiscent of the ground
state structure of the antiferromagnetic Ising model on a triangular
lattice, for which frustration effects lead to a highly degenerate
ground state. Finally, we observe 3 FLS's to crystallize into the
predicted groundstate but via an intermediate mesophase whose
stability remains unclear. An example of this multi-step
crystallization is illustrated in Figure~\ref{fig:E_vs_T}d. Note that the
mesophase is observed only on cooling of the liquid, and at
temperatures lower than the computed equilibrium transition between
liquid and crystal.

We also identified two scenarios for which the low-temperature state
was structurally and energetically different from the predicted ground
state.

ii) \emph{Metastable Polymorph Formation} ($14$ FLS) In these cases, a
recognizably crystalline structure was consistently obtained, albeit
with an energy strictly higher than the ground state. This scenario is
illustrated in Figure~\ref{fig:low_T}b, for which a complex crystalline
state with energy $-4/9$ and a huge cell size $\mathcal{Z}\geq 288$ was
obtained, while a much simpler and stabler crystalline structure with
energy $-1/2$ and cell size $\mathcal{Z}=16$ was available. In some
cases the crystal state formed can vary between different cooling
runs. This is clear evidence of polymorphism. We do not know, in each
case, whether the higher energy structure is, in fact, a true
equilibrium phase at the freezing temperature, favoured over the lower
energy structure by entropic effects, or a metastable crystal,
selected by the kinetics of ordering. The demonstration that our model
includes polymorphs is interesting given the prevalence of polymorphs
in molecular crystals. In one well studied case~\cite{roy}, a molecule
has been observed to crystallize into $7$ distinct crystal structures.

iii) \emph{Amorphous Solid Formation} Finally, in a single case, the
FLS \textbf{51c}, the liquid freezes with an apparent discontinuity of
the energy into an amorphous state with no recognizable trace of
crystallinity (see Figure~\ref{fig:low_T}c).  This transition is
accompanied by a dynamical arrest of the supercooled liquid, in the
quasi-critical scenario discussed in Section~\ref{sec:supercooling},
and the relaxation time scale becomes larger than our resolution of
$\tau = 10^3$ MC steps per site. On heating this state, negligible
hysteresis is observed. It is not clear whether this puzzling
behaviour corresponds to a conventional glass transition (for which
the energy would have to be continuous) or some sort of
polycrystalline aggregation. What is clear is that the \textbf{51c}
case represents the interesting possibility of the formation of an
amorphous solid stabilized entirely (by construction) by the same FLS
as stabilizes the crystal.

\section{On the Correlation between the FLS Structure and the
  Properties of the Liquid and Solids States}
\label{sec:stats}

In the previous sections, we have described the variety of behaviours
of our model, going through different aspects of its macroscopic
properties: crystalline ground states, liquids and supercooled
liquids, freezing transitions and low-temperature states. In this
Section, we examine a number of hypotheses associated with how the
choice of FLS determines the collective properties of the system. A
summary of our conclusions and the statistical data on which they are
based is presented in Table~\ref{tab:hypoth}.
 
 \begin{table*}[ht]
  \begin{center}
  \caption{ Summary of our correlation hypotheses and associated conclusions. Their validity is assessed by the $p$-value, corresponding to the likelihood of observing comparable correlations in the case of independent variables, as provided by either Pearson's analysis$^*$ or Student's $t$-test$^\dagger$. Numbers in parentheses are standard deviations. }
\label{tab:hypoth}
\includegraphics[width=\textwidth]{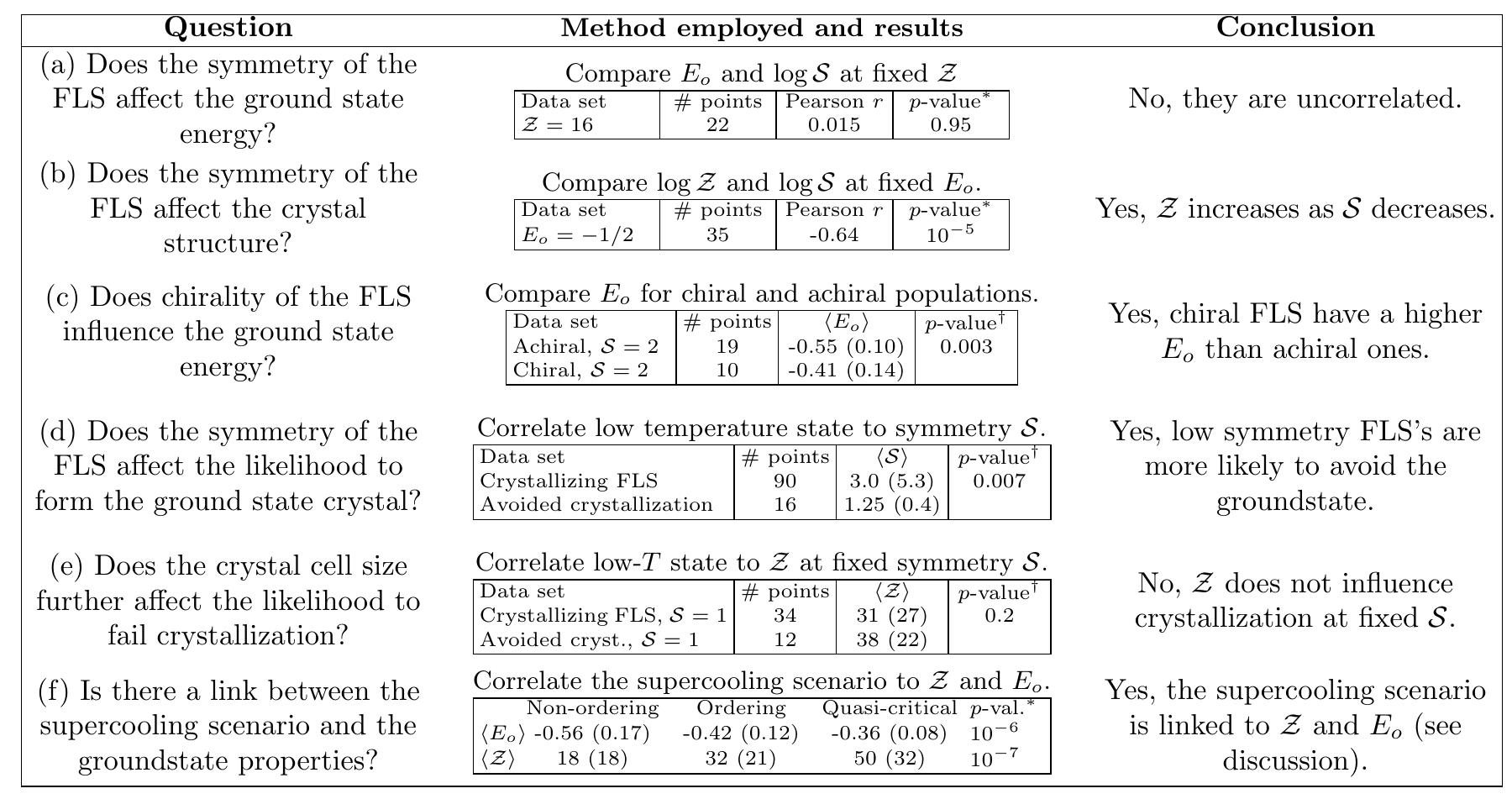}
  \end{center}
\end{table*}

\subsection{Structure of the Ground State}
\label{sec:SvsEo}

We start by considering the influence of the symmetry properties of
the FLS (the size $\mathcal{S}$ of the spatial point group of an FLS introduced in
Section~\ref{sec:model})on the structure of the ground state. In
Section~\ref{sec:GS} we introduced two observables to quantify the
latter: the ground state energy $E_o$ measures the quality of the
packing, while the cell size $\mathcal{Z}$ measures its complexity.
We disentangle the correlations between these quantities by
restricting the statistical analysis to sets of FLS for which one
parameter is fixed -- for instance, we analyze the correlation between
$\mathcal{S}$ and $E_o$ by considering the $22$ FLS's with crystal
cell $\mathcal{Z}=16$. This is made possible by the existence of sharp
peaks in the histograms of Figures~\ref{fig:Eo_and_Z}. Because
$\mathcal{Z}$ and $\mathcal{S}$ vary over several orders of magnitude,
we consider their logarithm when measuring correlations.

The connection between symmetry and ground state energy is
investigated in Table~\ref{tab:hypoth}a.  There appears to be no
significant correlation between $E_o$ and $\mathcal{S}$: the degree of
symmetry of the local structure does not influence its ground state
energy. This is perhaps surprising, as one might expect awkward
objects of low symmetry to have poor packing properties. We interpret
the independence between $E_o$ and $\mathcal{S}$ as the compensation
of two effects: on the one hand, high-symmetry objects are more likely
to fit well together; on the other hand, low-symmetry objects have
more rotational variants, thus offering a wider pool of combinations
to select the ground state from.

On the other hand, $\mathcal{S}$ and $\mathcal{Z}$ are significantly
correlated, with a correlation ratio $r = -0.64$, as reported in
Table~\ref{tab:hypoth}b. Thus the cell size increases with decreasing
symmetry. However, this trend is loose and admits exceptions: for
instance, highly symmetric FLS \textbf{7r}, for which $\mathcal{S}=6$,
has a large unit cell $\mathcal{Z}=64$ depicted in
Figure~\ref{fig:GS_pictures}(c). On the other hand, low-symmetry FLS
\textbf{32c} (with $\mathcal{S}=1$) crystallizes in a simple structure
with $\mathcal{Z}=7$, as depicted in Figure~\ref{fig:GS_pictures}(d).

\subsection{The special role of the chiral symmetry}
\label{sec:chiral_special}

As already hinted in Section~\ref{sec:Eo}, the ground state energies
of chiral FLS are generally higher than those of achiral
structures. This claim is properly confirmed in
Table~\ref{tab:hypoth}c, by comparing two groups of FLS with identical
degree of symmetry: on the one hand the 19 achiral FLS's with a plane
reflection symmetry only, and on the other hand the 10 chiral FLS's
with only a rotational symmetry of order 2. We observe a statistically
significant difference in the average ground state energy, while the
difference in crystal cell size is insignificant. This shows that
plane reflection symmetry of the structure plays a special role in
determining the ground state energy: chiral local structures pack less
densely than achiral structures. This special role of the chiral
symmetry in such discrete packing problems is, to our knowledge, an
original result of this study.

\subsection{Crystallization or not?}
\label{sec:crystallize}

We next turn to the question of the influence of structural properties
on the low temperature fate of the system. We consider on the one hand
the $90$ ``crystallizers'' which find their exact ground state, and on
the other the $16$ FLS's which do not find their ground state
structure, and form either a glassy state or a metastable
polymorph. We exclude defected crystals from this analysis, as they
lie in-between. The dominant contribution is the symmetry factor: low
symmetry strongly conditions towards avoided crystallization, as
demonstrated in Table~\ref{tab:hypoth}d.

Because low symmetry structures also tend to form large crystal cells
(Table~\ref{tab:hypoth}b), crystal formation also strongly correlates
with crystal cell size. However, we can refine this observation by
considering the set of FLS with $\mathcal{S}=1$. In
Table~\ref{tab:hypoth}e, we show that the value of $\mathcal{Z}$ does
not significantly affect the probability to fail crystallization
within this population.  To conclude, we have found that low symmetry
FLS tend to crystallize less than high symmetry ones, but that the
crystal cell size -- and hence its ``complexity'' -- is no further
impediment to crystallization. We note in particular that our three
``giant unit cell'' crystals with cell size $\mathcal{Z}>80$
straightforwardly freeze into a perfect crystalline structure.  On the
other hand a low ground state energy favours crystallization. The
latter observation can be intuitively explained by noting that a
crystalline structure that is a deep minimum of energy -- with low
$E_o$ -- is less likely to have viable competing polymorphs than a
high energy ground state.

Only a single FLS leads to an amorphous solid, as discussed in
Section~\ref{sec:lowT}, and we cannot make general statements on this
basis. We note however that it is chiral, with minimal symmetry
$\mathcal{S}=1$, has a high ground state energy $E_o = -3/10$, and a
large crystal cell $\mathcal{Z} = 80$.  Variants or extensions of the
present model could provide a larger population of similar amorphous
systems, and give a statistical basis to draw more general trends.

\subsection{Supercooling and ground state structure}
\label{sec:ZvsSupercooling}

The last point of our statistical analysis concerns the physical
properties of the supercooled liquid and its connection with
structural properties. In Section~\ref{sec:supercooling}, we
qualitatively distinguished three different behaviours in the
supercooled liquid: no trace of ordering, significant ordering, and
divergent time scale (``quasi-critical''). In Table~\ref{tab:hypoth}f
we correlate this classification with the ground state energies and
crystal cell sizes of all FLS's.

We observe a clear trend: high order in supercooled liquids correlates
with high ground state energy and large crystal cells. The first
observation is not surprising. Indeed, as we have seen in
Section~\ref{sec:liquids}, high ground state energy correspond to low
freezing temperatures, and thus to the possibility for the liquid to
overcome high entropic barriers to ordering. On the other hand, the
fact that crystal structure correlates with liquid ordering is rather
unexpected, as in a simple nucleation picture the liquid does not
``know'' about the crystal structure. This is not a symmetry effect:
restricting to the 26 chiral structures with $\mathcal{S}=1$, we
observe the same effect (although not resolved statistically).

\section{Discussion}
\label{sec:discussion}

What are the macroscopic consequences of local structure in liquids?
In this article, we addressed this question in the case when the
favoured local structures found in the liquid and in the solid are the
same -- that is, when the ground state corresponds to a dense
accumulation of a single FLS. We introduced a simple three-dimensional
lattice model in which the FLS is explicitly specified. This Favoured
Local Structures model on a 3D face-centered cubic lattice provides a
discrete set of $115$ such structures to choose from, and we reported
here on the comprehensive study of these systems. This large library
of structures is a unique feature of this work and allowed us to
employ an original statistical approach to correlate observable
features of the model and unfold underlying physical laws.  Given the
similarity between different instances of the model -- only the
geometry of the FLS varies -- the variety of observed behaviours is
noteworthy.

We focused on the structural properties of the ground states --
studying the relation between the symmetry of an FLS and its packing
properties -- and on the properties of the supercooled liquid --
addressing the questions of how much order can be accumulated prior to
freezing, and what solid phase will be formed at the freezing
transition. Our main results can be summarized as follows:
\begin{enumerate}[leftmargin=*]
\item All ground states are crystalline, with high variability in both
  the ground state energy $-1\le E_o \le -1/4$ and unit cell size
  $1\le \mathcal{Z}\le 128$.  This structural complexity does not
  appear to present any systematic kinetic obstacle to
  crystallization.
\item We found no evidence that low symmetry FLS's result in 'poor'
  (\emph{i.e.} high energy) crystals. Instead, we found that low
  symmetry FLS's tend to produce more complex crystals as
  characterised by larger unit cells. We interpreted this as resulting
  from compensating effects: low symmetry objects offer more
  orientational freedom than high symmetry ones, but the latter
  generally fit better together. We observed an exception to this rule
  in the case of chiral structures, which pack poorly, compared to
  achiral structures with equal degree of symmetry.
\item We observed only a single case of enantiomeric separation of a
  racemic mixture by crystallization. This exception corresponding to
  a chiral FLS that managed a particularly low energy groundstate.
\item The entropic cost of ordering in the liquid is so high in our
  model that we observe little organization prior to equilibrium
  crystallization. We regard this a consequence of the very strict
  definition of local structures in our model. Considering incomplete
  or multiple FLS could allow for more complex liquid behaviour.
\item Freezing transitions are typically strongly first-order and most
  liquids can be readily supercooled. Approaching the freezing
  transition, some liquids exhibit significant ordering and even
  apparently diverging relaxation times, while others remain
  unstructured. High ground state energy depresses the transition
  point towards low temperatures, allowing higher order in the
  liquid. More surprising, we found that systems whose crystalline
  structure is complex (i.e. large $\mathcal Z$) exhibit stronger
  ordering in the supercooled liquid, showing a significant
  correlation between the ground state structure and the amount of
  order in the amorphous state.
  \item Finally, we found that low symmetry and high ground state energy
  FLS tend to avoid crystallization by either falling into a
  metastable crystalline structure, or into an amorphous state. On the
  other hand the complexity of the crystalline structure does not
  affect the likelihood to crystallize. In particular, our giant unit
  cell crystals are readily formed in Monte-Carlo simulations of large
  systems.
\end{enumerate}

Perhaps the most telling conclusion from this study comes from a
negative result. Despite our comprehensive inclusion of all possible
local structures, we never found a local structure so geometrically
'awkward' that it lacked a crystalline groundstate: our enumerative
procedure to find crystalline structures always provided the lower
bound on observed energies. Furthermore, every choice of FLS (with a
single exception), exhibited a relatively rapid crystallization, if
not always to the ground state. While the choice of FLS can influence
the size of the unit cell in the crystal and the degree of ordering in
the supercooled liquid, we conclude that the existence and kinetic
accessibility of the crystalline phase are essentially independent of
the nature of the local structure. As for the liquid state, it is
clearly the multiplicity of the FLS, not its geometrical structure,
that exerts the most important influence with regards the accumulation
of structure in the liquid and any associated slowing down. This
conclusion challenges the long lived tradition of assigning
significance to the presence of specific local structures in low
temperature liquids.

\subsection*{Acknowledgements}

We thank Sinai Robins for pointing out the correspondence between
Hermite matrices and crystalline structures on a lattice. We thank
Hanna Gr\"onqvist for useful comments. PH acknowledges the financial
support of the Australian Research Council and PR acknowledges the
hospitality of the School of Chemistry at the University of Sydney
where this work was initiated.  Three dimensional pictures were
realized with Mayavi~\cite{mayavi}, plots with
Matplotlib~\cite{matplotlib}.

\appendix

\section{Enumerative method for determining ground states}
\label{sec:enumerative_method}

Our method for determining ground state structures and energies makes
use of a correspondence between crystalline base vectors on a lattice
and number-theoretical objects called Hermite matrices. A similar
approach was introduced by Hart and Forcade~\cite{hart} as a tool to
study derivative structures for metallic alloys. The idea is to list
all possible sets of crystalline base vectors (superlattice) on the
FCC lattice. There are $17602$ distinct superlattices of the FCC
lattice with up to $\mathcal{Z}=80$ sites per crystalline cell (having
eliminated redundancies due to reparametrization and lattice
symmetries). We use the correspondence between superlattices and
integer matrices in the Hermite normal form, completed by a reduction
using the FCC symmetries, to obtain this set of structures.  For each
such superlattice and every FLS, we find the best crystalline spin
configuration associated to this superlattice by simulated annealing
of the spin variables of a single crystal cell, with appropriate
boundary conditions. The small number of spins involved in this makes
us confident that the best spin structure is found. Whenever
necessary, we extend this procedure up to $128$ sites per cell. This
systematic approach is complemented with a more standard Monte-Carlo
simulated annealing using a rejection-free algorithm~\cite{bortz} of a
large system of up to $24^3$ sites, which always gives a resulting
minimal energy higher or equal to that obtained in the enumerative
procedure.  While these approaches can in principle fail and fall in
metastable minima, we are confident that most (if not all) of the
ground states are indeed the global ones.

\end{document}